\documentclass[a4paper,11pt]{article}

\usepackage{amsfonts}
\usepackage{algorithm}
\usepackage{algorithmic}

\newcommand{\RR}[2]{\mathbb{R}^{#1 \times #2}}

\newcommand{\rank}[1]{\mbox{rank}\left(#1\right)}

\newcommand{\ST}{\mathcal{S}}
\newcommand{\SVT}{\mathcal{D}}

\newcommand{\st}{\mbox{s.t.~}}

\newcommand{\refEq}[1]{Eq.(\ref{#1})}

\newcommand{\refAlg}[1]{Algorithm~\ref{#1}}

\def\marginset#1#2{                      

\setlength{\oddsidemargin}{#1}

\setlength{\evensidemargin}{0mm}

\setlength{\hoffset}{\paperwidth}

\addtolength{\hoffset}{-\oddsidemargin}

\addtolength{\hoffset}{-\textwidth}

\addtolength{\hoffset}{-\evensidemargin}

\setlength{\hoffset}{0.5\hoffset}

\addtolength{\hoffset}{-1in}

\setlength{\voffset}{-1in}

\setlength{\topmargin}{\paperheight}

\addtolength{\topmargin}{-\headheight}

\addtolength{\topmargin}{-\headsep}

\addtolength{\topmargin}{-\textheight}

\addtolength{\topmargin}{-\footskip}

\addtolength{\topmargin}{#2}

\setlength{\topmargin}{0.5\topmargin}
}

\marginset{0mm}{0mm}

\usepackage{times}
\usepackage{multirow}
\usepackage{fancyhdr}
\usepackage{threeparttable}
\usepackage{longtable}
\usepackage{color}

\usepackage{indentfirst}


\usepackage{verbatim}
\usepackage{url}
\usepackage[dvips]{graphicx}
\usepackage[fleqn]{amsmath}

\def\bfu{{\mathbf{u}}}
\def\bfv{{\mathbf{v}}}

\def\bfD{{\mathbf{D}}}
\def\bfE{{\mathbf{E}}}

\def\bfM{{\mathbf{M}}}

\def\bfX{{\mathbf{X}}}

\def\half{\frac{1}{2}~}

\begin{document}

\title{Exploring the genetic patterns of complex diseases via the integrative genome-wide approach}
\author{\\\\Ben Teng\\
Department of Computer Science and Institute of Theoretical and Computational Study,\\
 Hong Kong Baptist University\\\\\\
Can Yang\\
Department of Mathematics,\\ Hong Kong Baptist University\\\\\\
Jiming Liu\\
Department of Computer Science and Institute of Theoretical and Computational Study,\\ Hong Kong Baptist University\\\\\\
Zhipeng Cai\\
Department of Computer Science,\\ Georgia State University\\\\\\
Xiang Wan$^*$(Corresponding Author)\\Department of Computer Science and Institute of Theoretical and Computational Study,\\ Hong Kong Baptist University
\\*Email: xwan@comp.hkbu.edu.hk\\\\\\ }

\maketitle

\newpage

\begin{abstract}

Motivation: Genome-wide association studies (GWASs), which assay more than a million single nucleotide polymorphisms (SNPs) in thousands of individuals, have been widely used to
identify genetic risk variants for complex diseases. However, most of the variants that
have been identified contribute relatively small increments of risk and only explain a
small portion of the genetic variation in complex diseases. This is the so-called missing heritability problem. Evidence has indicated that many complex diseases are genetically related, meaning these diseases share common genetic risk variants. Therefore, exploring the genetic
correlations across multiple related studies could be a promising strategy for removing
spurious associations and identifying underlying genetic risk variants, and thereby uncovering the mystery of missing heritability in complex diseases.

Results: We present a general and robust method to identify genetic patterns from multiple large-scale genomic datasets. We treat the summary statistics as a matrix and demonstrate that genetic patterns will form a low-rank matrix plus a sparse component. Hence, we formulate the problem as a matrix recovering problem, where we aim to discover risk variants shared by multiple diseases/traits and those for each individual disease/trait. We propose a convex formulation for matrix recovery and an efficient algorithm to solve the problem. We demonstrate the advantages of our method using both synthesized datasets and real datasets. The experimental results show that our method can successfully reconstruct both the shared and the individual genetic patterns from summary statistics and achieve better performance compared with alternative methods under a wide range of scenarios.

Availability: The MATLAB code is available at:\url{http://www.comp.hkbu.edu.hk/~xwan/low_rank.zip}.

\end{abstract}

\section{Introduction}
Many common human diseases, such as type-1 and type-2 diabetes, depression, schizophrenia, and prostate cancer, are influenced by several genetic and environmental factors. Scientists and public health officials have great interests to find genetic patterns associated with complex diseases, not only to advance our understanding of multi-gene disorders, but also to provide more insights into complex diseases. Disease association studies have provided substantial evidence for supporting that complex diseases originate in disorders of multiple genes \cite{mcclellan2007schizophrenia,morris2012large}. Nevertheless, until recently the full-coverage identification of the genetic variants contributing to complex diseases has been staggering and difficult.

After the completion of the Human Genome Project \cite{venter2001sequence, lander2001initial} and the initiation of the International HapMap Project \cite{sachidanandam2001map}, interest has focused on genome-wide association studies (GWASs), in which the goal is to identify single-nucleotide polymorphisms (SNPs) that are associated with complex diseases (such as diabetes) or traits (such as human height). As of Dec. 2014, more than $15,000$ SNPs have been reported to be associated with at least one disease/trait at the genome-wide significance level ($P$-value$\le 5 \times 10^{-8}$) \cite{Hindorff2015}. However, most of the findings only explain a small portion of the genetic contributions to complex diseases. For example, all of the $18$ SNPs identified in type 2 diabetes (T2D) only account for about 6\% of the inherited risk \cite{manolio2009finding}. There is still a large portion of disease/trait heritability that remains unexplained. This is the so-called missing heritability problem \cite{manolio2009finding,maher2008personal}, which is often used to denote the gap between the expected heritability of many common diseases, as estimated by family and twin studies, and the overall additive heritability obtained by accumulating the effects of all of the SNPs that have been found to be significantly associated with these conditions.

A recent study \cite{yang2010common} has suggested that most of the heritability is not missing but can be explained by the effects of many genetic variants, with each variant probably contributing a weak effect. However, finding variants with small effects is very challenging in computation because the traditional single-locus based test cannot identify such variants and the number of groups of multiple variants to be investigated in GWAS is astronomical. In addition, in the high-dimensional and low-sample size settings of GWAS, many irrelevant variants tend to have high sample correlations due to randomness, which makes GWAS prone to false scientific discoveries. To solve the missing heritability problem, the large sample size is required, but such a requirement is usually beyond the
capacity of a single GWAS, as the sample recruitment is expensive and time consuming.

Evidence has indicated that many complex diseases are genetically related \cite{sivakumaran2011abundant,vattikuti2012heritability,cross2013genetic,cross2013identification}, meaning that these diseases share common genetic risk variants. This suggests that an integrative analysis of related genomic data could be a promising strategy for removing
spurious associations and identifying risk genetic variants with small effects, and thus
finding the missing heritability of complex diseases. As high-throughput data
acquisition becomes popular in biomedical research, new computational methods for
large-scale data analysis become more and more important.

When analyzing genomic data from multiple related studies, the ideal scenario is for the individual-level data to be available for all of
the included studies, but this may be difficult to achieve due to restrictions on sharing
individual-level data. In fact, summary data (mostly $P$-values or $z$-scores) are more
frequently shared. To identify significant SNPs shared by all of the included studies, the commonly used statistical approach is to combine $P$-values
 using Fisher's method \cite{fisher1934statistical}.  \cite{goods1955weighted} generalized Fisher's method to include weights when combining $P$-values. \cite{stouffer1949american} suggested using the inverse normal transformation and Mosteller and \cite{mosteller1970selected} further generalized Stouffer's method by including weight when combining $z$-scores. There are two issues in such traditional statistical approaches. First, one small $P$-value can overwhelm many large $P$-values and dominate the test statistic. In a high-dimensional and low-sample size settings, many irrelevant variants tend to have high significance due to randomness, which may cause wrong statistical inferences.
Second, the information about genetic correlations between SNPs in the original data is completely lost after combining $P$-values.  This information is necessary for understanding the genetic architecture of complex diseases because common complex diseases are associated with multiple genetic variants.

To identify shared genetic structures across multiple related studies, one feasible approach is to conduct a biclustering analysis on a matrix of summary statistics, in which the rows represent studies and the columns represent genetic variants, to simultaneously group studies and genetic variants. Many biclustering methods have been proposed and some comprehensive reviews of biclustering methods can be found in  \cite{madeira2004biclustering}, \cite{prelic2006systematic}, and \cite{busygin2008biclustering}. However, the traditional biclustering methods do not perform well on genomic data because genomic data is high dimensional and its most genetic variants are irrelevant. To obtain sparse and interpretable biclusters, a novel statistical approach, \textbf{sparseBC}, is recently proposed, which adopts an $l_1$ penalty to the means of the biclusters \cite{tan2013sparse}. A big drawback of \textbf{sparseBC} is that  it does not allow for overlapping biclusters, which limits its application in genomic data analysis because the shared genetic patterns in GWASs may be very complex. Furthermore, in genomic data, besides the shared genetic structure, each disease/trait owns some distinct genetic variants. The typical biclustering model may treat them as noises and discard them.

In this paper, we introduce a new method to identify genetic patterns in high dimensional genomic data. Our method possesses several advantages over existing works. First, our method admits a single model to detect both shared and individual genetic patterns among multiple studies. Second, our method employs two tuning parameters that control the size of the shared genetic pattern and the numbers of individual signals. The choices of these parameters have the solid theoretical support. Third, our method produces the unique global minimizer to a convex problem, which means that the solution is always stable.



To demonstrate the performance of our proposed method, we conduct
comparison experiments using both synthesized datasets and real datasets.
Simulation results show that the proposed method outperforms existing
methods in many settings. A large dataset containing 32 GWASs is
also analyzed to demonstrate the advantage of our method. Specifically,
we propose the convex formulation, the algorithm, and the parameter selection in Section
2. Simulation studies and real data analysis are presented in
Section 3. We conclude the paper with some discussions in
Section 4.

\section{Methods}

\subsection{Formulation}
Mathematically, the summary statistics from multiple related studies
 can be expressed as a matrix $\bfD\in\RR{n}{p}$, where each entry $d_{ij}$ is a $z$-score (if only $P$-values are available, we can transform them into $z$-scores), and $n$ and $p$ are the numbers of studies and SNPs, respectively. Our goal is to (1) detect shared genetic patterns across studies, which can be represented as sparse biclusters in this matrix and (2) detect individual genetic variants for each study, which we assume are randomly distributed and sparse.
 Since the sparsity of biclusters in a matrix indicates a low-rank property (please see examples in simulation studies), the problem of identifying these two types of genetic patterns can be treated as a problem of recovering a low-rank component $\bfX$ and a sparse component $\bfE$ from the input data $\bfD$. Our proposed approach is based on the assumed sparsity of genetic patterns because in large-scale genomic data, most genetic variants are irrelevant.

We propose to use the following decomposition model to detect genetic patterns from noisy input:
\begin{align}\label{eq:decomp}
\bfD = \bfX + \bfE + \epsilon,
\end{align}
where $\bfX$ is a low-rank component, $\bfE$ is a sparse component, and $\epsilon$ is a noise component. In GWAS data analysis, the low-rank component corresponds to the causal SNPs that are shared by several diseases/traits. The sparse component corresponds to the causal SNPs that affect one specific disease/trait. The noise component corresponds to the measurement error, which is often modeled by i.i.d. Gaussian distribution with a zero mean.

Naturally, to achieve the decomposition, the following minimization problem is considered:
\begin{align}\label{eq:nonrelax}
\min_{\bfX,\bfE,\epsilon} ~ &{\half}\|\epsilon\|_F^2 + \alpha \rank{\bfX} + \beta \|\bfE\|_0 \nonumber \\
\st ~ &\bfD = \bfX + \bfE + \epsilon,
\end{align}
where $\|\epsilon\|_F=\sqrt{\sum_{i,j}{\epsilon_{ij}^2}}$ is the Frobenious norm and $\|\bfE\|_0$ is the $\ell_0$-norm that counts the number of nonzero values in $\bfE$. The solution to \refEq{eq:nonrelax} will give a penalized maximum likelihood estimate with respect to the variables $\bfX,\bfE,\epsilon$.

However, the proposed model in \refEq{eq:nonrelax} is intractable and NP-hard. Thus, in order to effectively recover $\bfX$ and $\bfE$, we use the convex relaxation to replace the $\rank{\cdot}$ by the nuclear norm and the $\ell_0$-norm by the $\ell_1$-norm. Here, the nuclear norm is defined as $\|\bfX\|_{*}=\sum_{i=1}^r{\sigma_i}$, where $\sigma_1,\cdots,\sigma_r$ are the singular values of $\bfX$. It is the tightest convex surrogate to the rank operator \cite{fazel2002matrix} and has been widely used for low-rank matrix recovery \cite{candes2011robust}. The $\ell_1$-norm is defined as $\|\bfX\|_{1}=\sum_{i,j}{|X_{ij}|}$. The $\ell_1$ relaxation has proven to be a powerful technique for sparse signal recovery \cite{tropp2006just}.

Finally, instead of directly solving \refEq{eq:nonrelax}, we solve the following problem,
{\footnotesize
\begin{align}\label{eq:rpla}
   \mathcal{F}(X,E) = \min_{\bfX,\bfE}{~~\frac{1}{2}\|\bfD-\bfX-\bfE\|_F^2+\alpha\|\bfX\|_*+\beta\|\bfE\|_1}.
\end{align}}

It is easy to prove that \refEq{eq:rpla} is a convex problem and therefore, the global optimal solution is unique. We will introduce the algorithm to solve this optimization problem in the next subsection.

\subsection{Algorithm}\label{sec:alg}

The optimization problem of \refEq{eq:rpla} can be solved by alternatively solving the following two sub-problems until convergence:
\begin{align}
\hat\bfX & \leftarrow \arg\min_{\bfX}\mathcal{F}(\bfX,\hat\bfE) \label{eq:admm_1}\\
\hat\bfE & \leftarrow \arg\min_{\bfE}\mathcal{F}(\hat\bfX,\bfE) \label{eq:admm_2}.
\end{align}

The theoretical proof for the convergence can be found in \cite{boyd2010distributed}.

The problem in \refEq{eq:admm_1} can be reduced to
{\small
\begin{align} \label{eq:B-step}
\min_{\bfX}{~~\frac{1}{2}\|\bfD-\hat\bfE-\bfX\|_F^2+\alpha\|\bfX\|_*},
\end{align}}
which becomes a nuclear-norm regularized least-squares problem and has the following closed-form solution \cite{cai2010singular},
\begin{align}\label{eq:svt_solution}
\hat\bfX = \SVT_{\alpha}\left(\bfD-\hat\bfE\right),
\end{align}
where $\SVT_{\lambda}$ refers to the singular value thresholding (SVT)
\begin{align}\label{eq:svt}
\SVT_{\lambda}(\bfM) = \sum_{i=1}^{r}{(\sigma_i-\lambda)_+\bfu_i\bfv_i^T}.
\end{align}
Here, $(x)_+ = \max(x,0)$. $\{\bfu_i\}$, $\{\bfv_i\}$, and $\{\sigma_i\}$ are the left singular vectors, the right singular vectors, and the singular values of $\bfM$, respectively.

The problem in \refEq{eq:admm_2} can be rewritten as
\begin{align}\label{eq:E-step}
\min_{\bfE}\frac{1}{2}\|\bfD-\hat\bfX-\bfE\|_F^2+\beta\|\bfE\|_1.
\end{align}
It admits a closed-form solution
\begin{align}
\hat\bfE = \ST_{\beta}\left(\bfD-\hat\bfX\right),
\end{align}
where $\ST_{\beta}(\bfM)_{ij}=\mbox{sign}(M_{ij})(M_{ij}-\beta)_+$ refers to the elementwise soft-thresholding operator \cite{boyd2010distributed}.

Overall, the algorithm to optimize the proposed model in \refEq{eq:rpla} is summarized in \refAlg{alg:rpla}. It will give a global optimal solution independent of initialization.

\begin{algorithm}
    \caption{The algorithm to solve \refEq{eq:rpla}.}\label{alg:rpla}
    \begin{algorithmic}[1]
    \algsetup{linenodelimiter=.}
        \STATE {\bf Input:} $\bfD$
        \STATE Initialize all variables to be zero.
        \REPEAT
            \STATE Update $\bfX$ by solving \refEq{eq:B-step} via singular value thresholding.
            \STATE Update $\bfE$ by solving \refEq{eq:E-step} via soft thresholding.
        \UNTIL{convergence}
        \STATE {\bf Output:} $\hat{\bfX}$ and $\hat{\bfE}$
    \end{algorithmic}
\end{algorithm}

\subsection{Parameter selection}

There are two parameters in our model, which can be estimated properly via the analysis of the size of the input matrix $(n,p)$ and the standard variation of the noise $\sigma$ \cite{candes2011robust,zhou2010stable}.

The relative weight $\lambda=\beta/\alpha$ balances the two terms in $\alpha\|\bfX\|_*+\beta\|\bfE\|_1$ and consequently controls the rank of $\bfX$ and the sparsity of $\bfE$. \cite{candes2011robust} has proved that $\lambda=1/\sqrt{m}$ gives a large probability of recovering $\bfX$ and $\bfE$ under their assumed conditions and stated that this value can be adjusted slightly to obtain the best results in specific applications. Here, $m$ is the larger dimension of the input matrix. In our problem, $m=p$, i.e. the number of SNPs. However, on real datasets, the shared SNPs rarely form a perfectly low-rank matrix, and we use $\beta=2\alpha/\sqrt{p}$ to keep sufficient variations in $\bfX$.

The parameter $\alpha$ serves as a threshold in the SVT step in \refEq{eq:svt}. It should be large enough to threshold out the noise but not too large to over-shrink the signal \cite{zhou2010stable}. A proper value is $\alpha=(\sqrt{n}+\sqrt{p})\sigma$, which is the expected $\ell_2$-norm of a $n\times p$ random matrix with entries sampled from $\mathcal{N}(0,\sigma^2)$. As SNPs are sparse in the data, we can estimate $\sigma$ from the data by the median-absolute-deviation estimator \cite{meer1991robust}
\begin{align}
\hat\sigma = 1.48~\mbox{median}\left\{|\bfD-\mbox{median}(\bfD)|\right\}.
\end{align}

\section{RESULTS}

\subsection{Simulation studies}

We first compare the performance of our method under four simulation studies, with three existing biclustering methods: sparseBC (sparse biclustering) \cite{tan2013sparse}, LAS \cite{shabalin2009finding} and SSVD \cite{lee2010biclustering}. Since biclustering methods search for sample-variable associations in the form of distinguished submatrices of the data matrix, we consider the entry $(i,j)$ that belongs to one of the resulting biclusters which meet a predefined criterion as the reported association. Specially, for sparse biclustering method, we use the parameters that have been mentioned in \cite{tan2013sparse}, and the entries in the clusters which satisfy a preselected cutoff are recognized as the final result. For LAS, we use the default settings. For SSVD that uses a variant of singular value decomposition to find biclusters, we try different setting of parameters and report the best one as its  result. LAS and SSVD can detect overlapping biclusters but sometimes they report the entire matrix as one bicluster. Thus, for both LAS and SSVD, the biclusters that contain the entire matrix are discarded. For our method, the parameters are selected as stated in Section 2. Then we use a threshold $T$ to determine whether the entries $(i,j)$ of matrix is reported as the result or not by comparing the value of $X(i,j)$ and $E(i,j)$ with $T$.

We evaluate each method in the term of $F1$-score, which can be calculated as following:

\begin{align}
precision = \frac{tp}{tp+fp},
\end{align}
\begin{align}
recall = \frac{tp}{tp+fn},
\end{align}
\begin{align}
F\text{1-score} = \frac{2*precision*recall}{precision+recall},
\end{align}
where $tp$ and $fp$ denote the number of true positives and false positives, respectively, and $fn$ denotes the number of false negatives.

\subsubsection{Simulation settings}

\begin{figure}
  \centering
  \includegraphics[width=1 \linewidth]{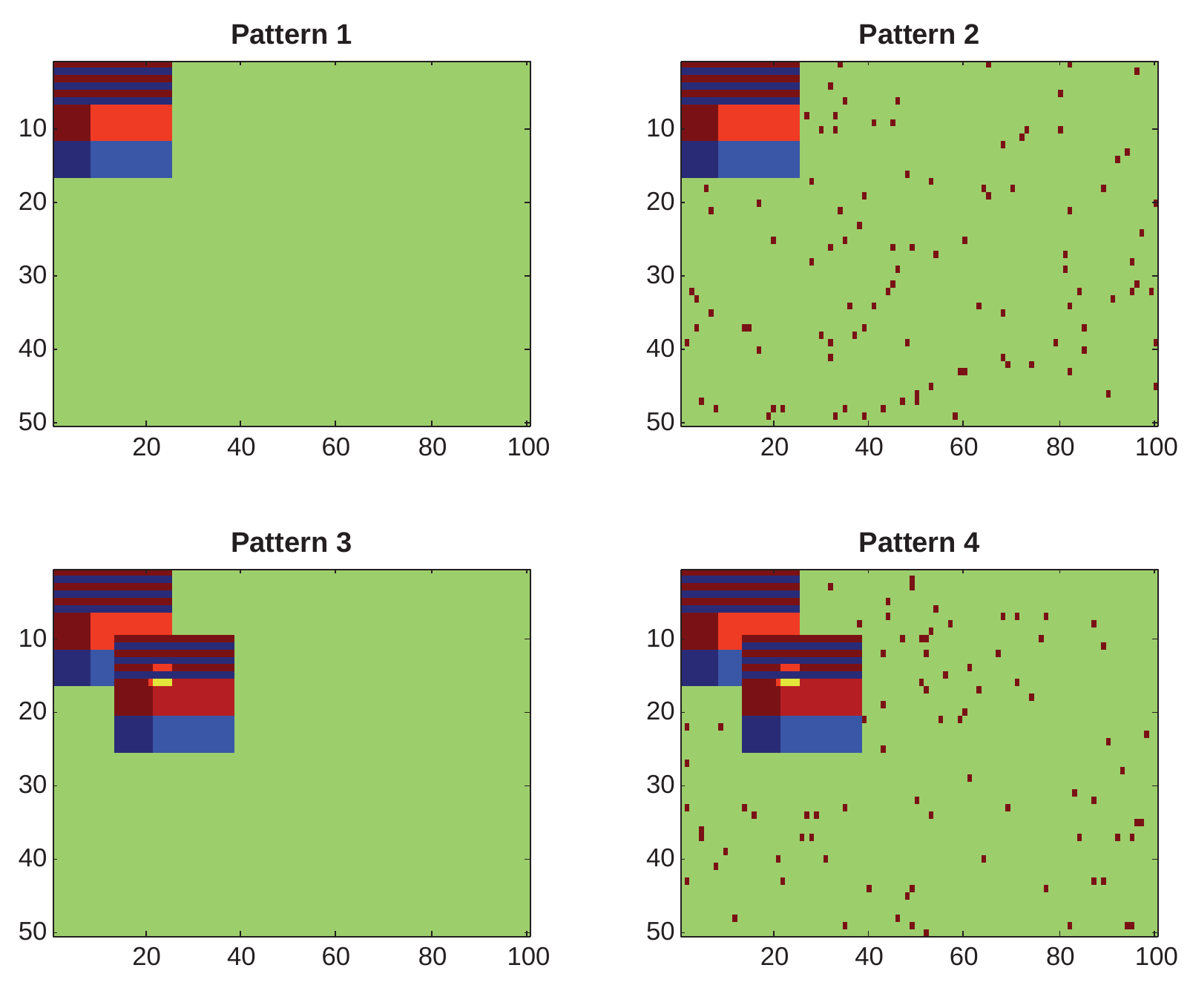}
  \caption{Four scenarios in our simulation study. Pattern 1 contains a rank-1 component representing one bicluster. Pattern 2 adds some sparse signals in Pattern 1. Pattern 3 contains a rank-2 component representing two overlapped biclusters. Pattern 4 contains sparse signal in addition to overlapped biclusters. }\label{fig:settings}
\end{figure}

We adopt four patterns (each in one simulation study) illustrated in Figure \ref{fig:settings} to generate synthetic data.
\begin{itemize}

\item Pattern 1 adopts a case from \cite{lee2010biclustering}, which generated a rank-1 true signal matrix. Let $\bfM = d\bf{u_1}\bf{v_1^T}$ be a $100 \times 50$ matrix with
 $d =50$, $\hat{v_1}=[10,9,8,7,6,5,4,3,r(2,17),r(0,75)]$, $\hat{u_1}=[10,-10,8,$ $-8,5,-5,r(-3,5),r(0,34)]^T$, $u_1=\hat{u_1}/\|\hat{u_1}\|_2$, and $v_1=\hat{v_1}/\|\hat{v_1}\|_2$, where $r(a,b)$ denotes a vector of length $b$ with all entries equal $a$.
  This case simulates the shared causal SNPs among several studies.

\item  Pattern 2 extends Pattern 1 by adding some sparse signals. That is, we generate a sparse component $\bfE$, whose entries are independently distributed, each taking on value 0 with probability $1-p_s$, and value 6 with probability $p_s = 0.01$.

\item  Pattern 3 adopts the case from \cite{tan2013sparse}, which generated two overlapping biclusters. Let $\bfM = d(\bf{u_1}\bf{v_1^T} +   \bf{u_2}\bf{v_2^T})$ be a $100 \times 50$ matrix with $d =50$, $u_1$ and $v_1$ as defined in simulation 1, $\hat{u_2}=[r(0,13),10,9,8,7,\\6,5,4,3,r(2,17),r(0,62)]$, $\hat{v_2}=[r(0,9),$ $10,-9,8,-7,6,-5,r(4,5),r(-3,5),r(0,25)]^T$, $u_2=\hat{u_2}/\|\hat{u_2}\|_2$, and $v_2=\hat{v_2}/\|\hat{v_2}\|_2$.

\item  Pattern 4 extends Pattern 3 by adding some sparse signals following the same way as Pattern 2.

\end{itemize}

\subsubsection{Data generation}
\begin{figure}
  \centering
  \includegraphics[width=1 \linewidth]{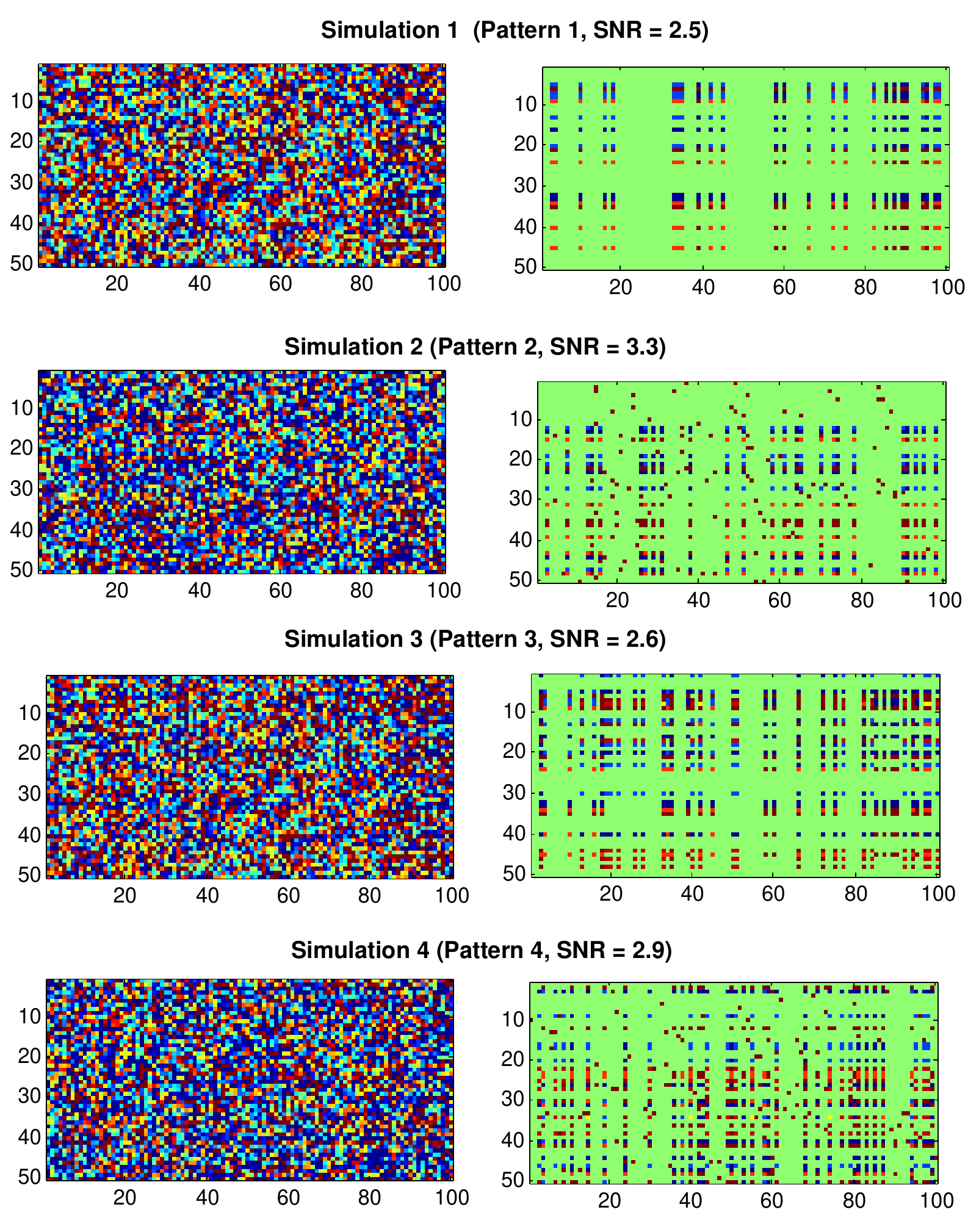}
  \caption{Illustrations of four simulations. For each simulation, the generated matrix with noises is shown in the left panel and the groundtruth matrix is shown in the right panel. In the groundtruth matrix, the red entries indicate the true signals.}\label{fig:data}
\end{figure}

Given a specific pattern mentioned above, we first generate the data matrix. To simulate the real situation, we randomly shuffle the rows and the columns. Next, we add Gaussian noise $\epsilon \sim \mathcal{N}(0,1)$ to each item. Figure \ref{fig:data} illustrates the groundtruth data and the generated data. For each generated data matrix, we also compute the signal to noise ratio (SNR). To illustrate how the methods perform for the data with different SNRs, we further scale down the ground true signal by dividing the original values by 1.2 and 1.5, respectively.

\subsubsection{Simulation results}

\begin{figure}
  \centering
  \includegraphics[width=1 \linewidth]{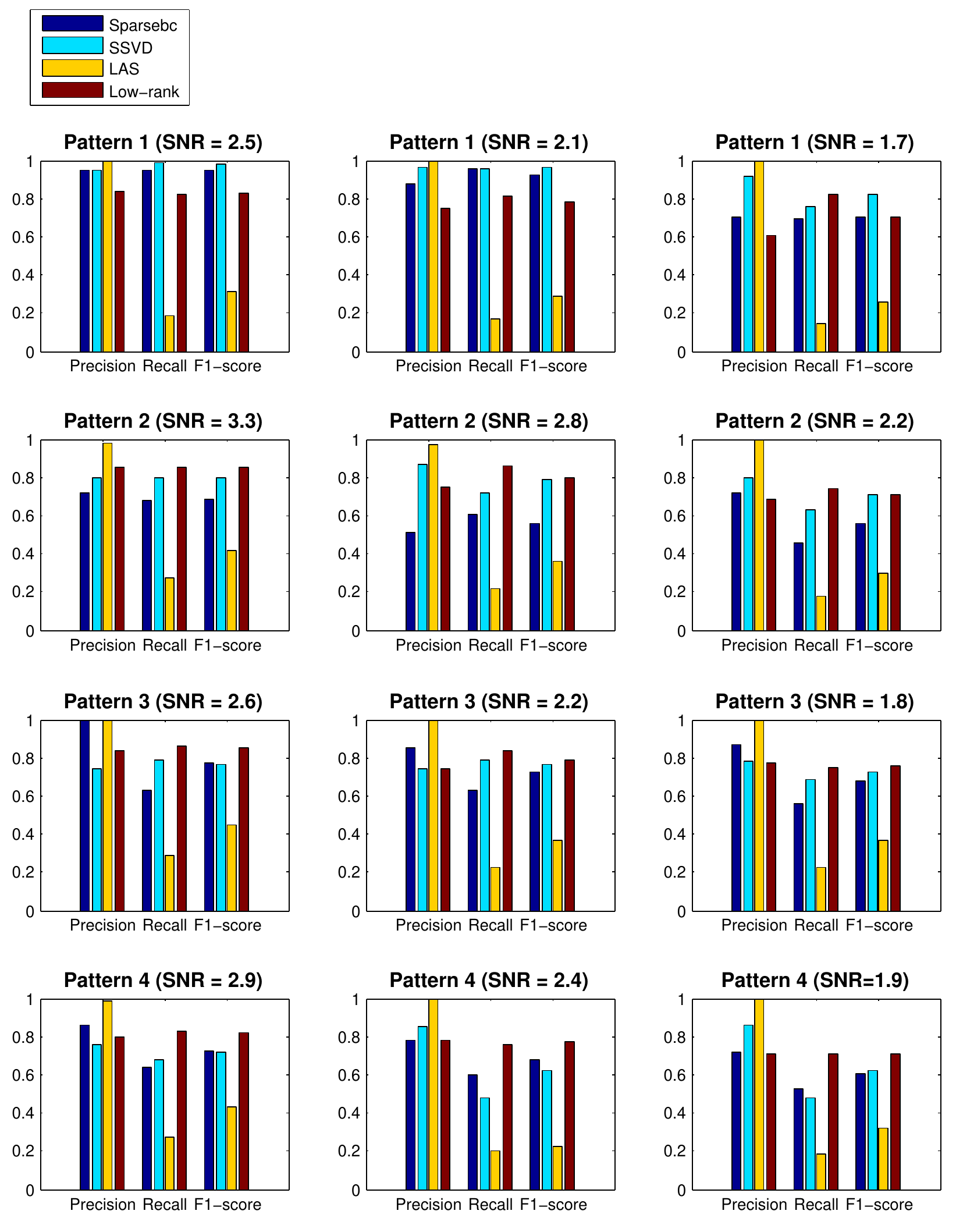}
  \caption{Comparison results of different methods in four simulation studies, each using one pre-defined pattern. }\label{fig:results}
\end{figure}

\begin{figure}
  \centering
  \includegraphics[width=1.05 \linewidth]{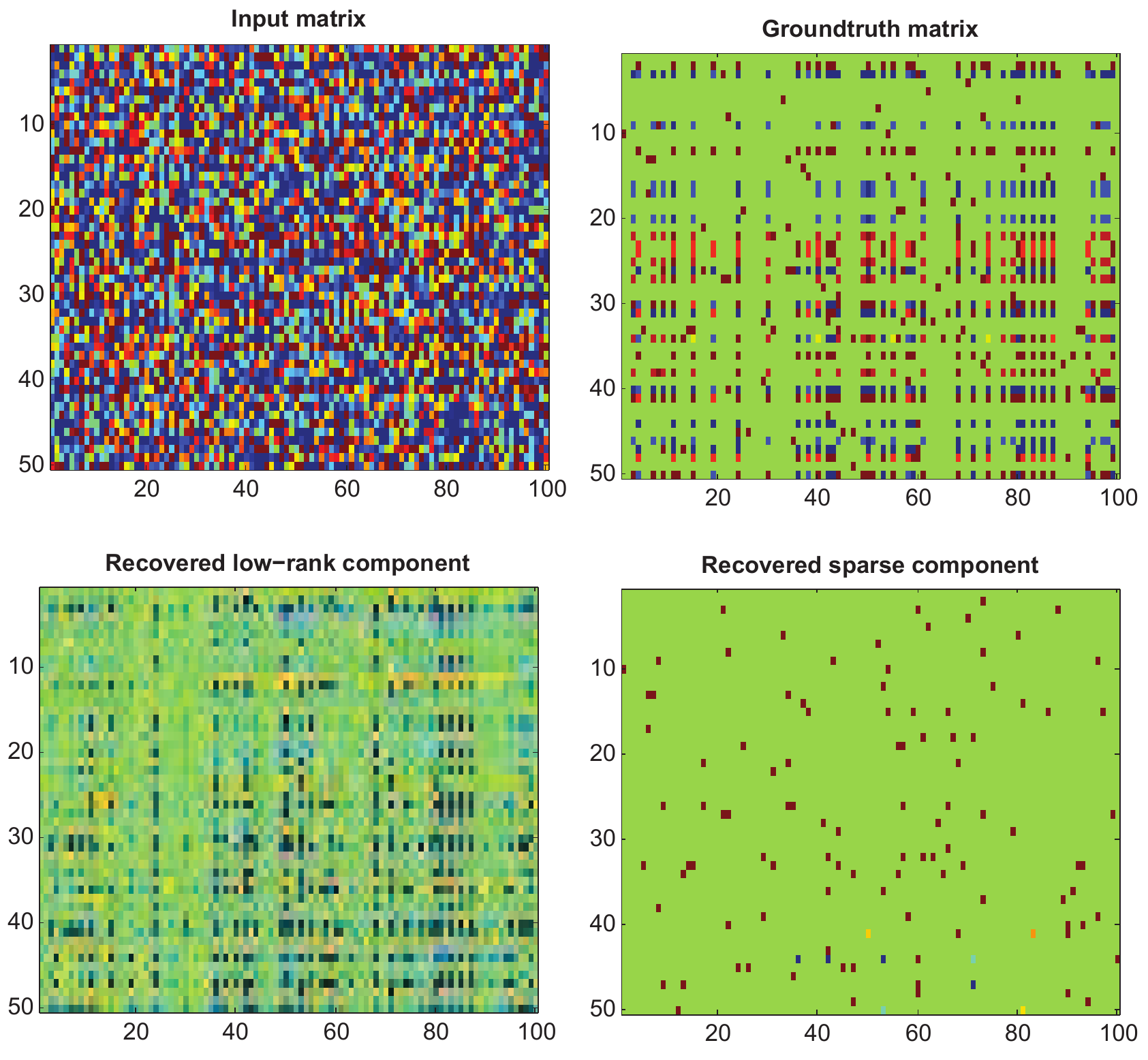}
  \caption{An illustration of the simulation result. The low-rank component and the sparse component are recovered by our method.}\label{fig:sim4_result}
\end{figure}

The results of four simulation studies are shown in Figure \ref{fig:results}. We use `low-rank' to represent our method as our model is to find biclusters via a low-rank approximation. The details of the simulation results can be found in the supplementary materials.
In general, our proposed method achieves comparable performance in the first and third simulation studies and performs better than other three  methods in the second and fourth simulation studies. This is because the classical biclustering methods suffer from several limitations, such as missing some entries for overlapped biclusters and the inability to identify the disease/trait-specific entries. Figure \ref{fig:sim4_result} shows one result in the fourth pattern. Our proposed method can successfully recover a low-rank component and a sparse component from raw data. In the first simulation, the $F1$-scores of sparse biclustering method and SSVD method almost get to 1. The reason why our method performs worse is that we use the default parameters which are not best fit for this simulation set-up. When adjusting the parameters, our method can also get a high $F1$-score.
Furthermore, we can observe from Figure \ref{fig:results} that our method always perform equally well in terms of both precision and recall while the other three methods often favor precision against recall. In the large-scale data analysis, the conservative method with high precision and low recall may not be suitable for new discoveries because most signals are irrelevant. For such situations, our method has a clear advantage over competitors.

\subsection{Real application}

We applied our method to analyze 32 independent diseases/traits, including

\begin{itemize}

\item 3 anthropometrics related data.

\item 9 pyschiatry related data.

\item 8 CAD data.

\item 2 social science studies

\item 2 glycaemic traits

\item 7 inflammatory bowel disease data.
\item  systemic lupus erythematosus
\item  parkinson
\end{itemize}

The details of the data sets including the references and the web link for downloading the data can be found in the supplementary materials.
Since each study reports different SNPs, we take the SNPs that are reported by at least 28 diseases/traits and obtain their $P$-values and impute the missing ones. Finally, we get a $P$-value matrix $P \in R^{466423 \times 32}$ for these 32 diseases/traits. Next, we convert the $P$-value matrix to the $z$-score matrix $Z\in R^{466423 \times 32}$.  We analyze this data set using our method on a desktop PC with 2.40GHz CPU and 4GB RAM. The running time of our method on 32 GWASs data sets is only 152.1s. The three alterative methods investigated in this work cannot be applied due to the large size of the data.

\begin{figure}
  \centering
   \includegraphics[width=1.05 \linewidth]{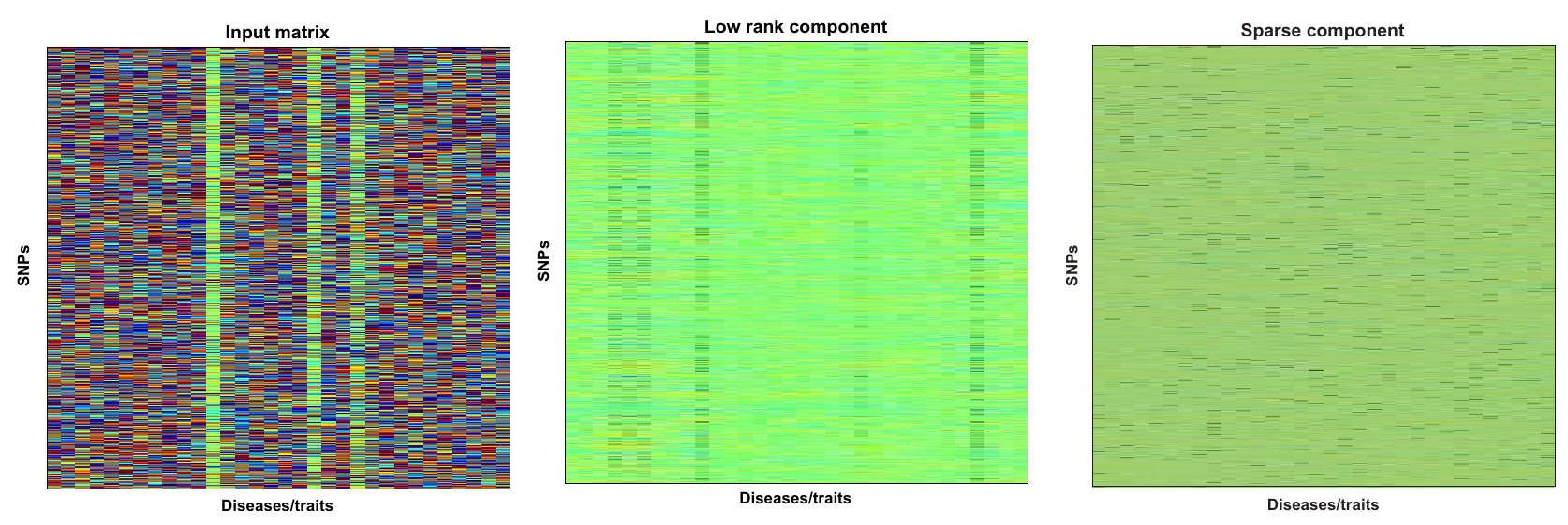}

  \caption{The experiment results on 32 GWASs. The low-rank component (middle panel) and the sparse component (right panel) are recovered by our method.}\label{fig:real_result}

\end{figure}

\begin{figure}
  \centering
   \includegraphics[width=1.05 \linewidth]{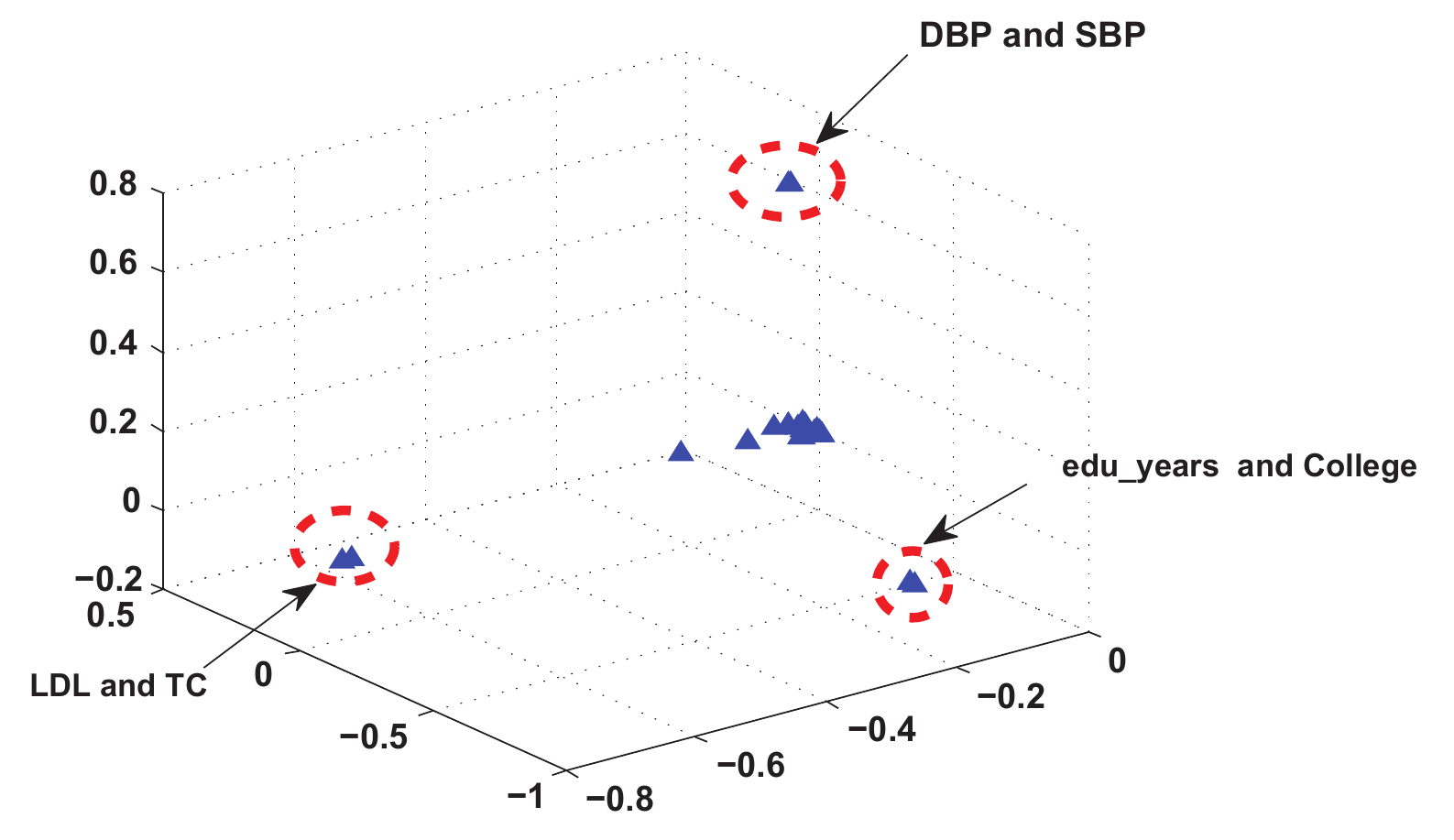}

  \caption{The geometric relationships of all studies using the coordinates derived from the first three right singular vectors of the recovered low-rank matrix.}\label{fig:low_zoom}

\end{figure}

The experiment results are given in Figure \ref{fig:real_result}. The shared causal SNPs are presented in the low-rank component and individual-specific SNPs are shown in the sparse component. We take the first three right singular vectors of the recovered low-rank matrix and use them as the coordinate of each study in Figure \ref{fig:low_zoom}. From Figure \ref{fig:low_zoom}, it is clear to see that 3 clusters are recovered from 32 diseases/traits:
\begin{itemize}
\item 2 social science studies (edu\_years and college);
\item diastolic blood pressure and systolic blood pressure (DBP and SBP);
\item total cholesterol and low density lipoprotein (TC and LDL).
\end{itemize}

The diseases/traits in each cluster are highly relevant with each other. We compare the identified causal SNPs by our method on 32 GWAS data with some previous findings. For 3 pairs of diseases/traits that are clustered together, we mainly investigate the shared SNPs that are identified by our method.
For two social science related data, our method has detected SNP \emph{rs3789044}, SNP \emph{rs12046747}, and SNP \emph{rs12853561}, which are mapped to genes \emph{LRRN2} and \emph{STK24}, respectively. These were reported in the original article \cite{rietveld2013gwas} because they have significant $P$-values (the details are provided in the supplementary materials).  However, besides those SNPs with significant $P$-values, our method has also identified some  locus with moderate $P$-values. SNP \emph{rs2532269}, whose original $P$-values are $1.01\times 10{^{-4}}$ in edu\_years data and $1.11\times 10{^{-4}}$ in college data, is detected as a causal SNP by our method. This SNP was previously reported ($P$-value $= 2 \times 10{^{-11}}$) \cite{early2012common} and mapped to the gene \emph{KIAA1267}. This gene is highly connected with Koolen-De Vries syndrome. Koolen-De Vries syndrome is characterized by moderate to severe intellectual disability, hypotonia, friendly demeanor, and highly distinctive facial features, including tall, broad forehead, long face, upslanting palpebral fissures, epicanthal folds, tubular nose with bulbous nasal tip, and large ears \cite{koolen2012mutations}.

For diastolic blood pressure and systolic blood pressure, the identified SNPs in our experiment are also connected with some previously published genes, such as \emph{ULK4, FGF5} and \emph{C10orf107} \cite{international2011genetic}. Similarly, some additional locus are identified by the low-rank component. SNP \emph{rs4986172} (original $P$-values in SBP data and DBP data are $3.09 \times 10^{-5}$ and $0.0172$, respectively), located in the gene \emph{ACBD4}, is detected by the low-rank component. This gene has been associated with high blood pressure in \cite{newton2009eight}.

To illustrate the power of our method in identifying the causal SNPs that do not shared by several diseases/traits, we take the result of bipolar disorder as an example. The SNPs in the result of bipolar disorder can be matched to \emph{ANK3, CACNA1C, SYNE1} and \emph{PBRM1}, which have been confirmed to be associated with bipolar disorder \cite{chung2014gpa}. The detailed results of other diseases/traits can be found in the supplementary materials.
Clearly, the experiment results show that not only can our method recognize SNPs with small $P$-values, but also detect those SNPs with moderate or weak $P$-values.

\newpage

%



\section{Discussion}
Finding weak-effect variants to explain the missing heritability of complex diseases is a challenging task and bottlenecked by the available sample size of GWAS. Based on the fact that related diseases/traits tend to co-occur, discovering shared genetic components among related studies becomes a popular way to address this issue. In the last few years, hundreds of GWASs have been carried out. Therefore, it is timely to systematically investigate GWAS data sets to find those shared patterns for comprehensive understanding of the genetic architecture of complex diseases/traits. In this work, we present a novel method for exploring the genetic patterns of complex diseases. We assume that causal SNPs can be divided into two categories: SNPs shared by multiple diseases/traits and SNPs for individual disease/trait. Thus, by modeling the problem as recovering a low-rank component and a sparse component from a noise matrix, we formulate it as a convex optimization problem. To demonstrate the performance of our proposed method, we conduct several simulation studies under different settings. Simulation results show that the proposed method outperforms three alternative methods in many settings. In the real data studies, we collect 32 large-scale GWAS data sets. We have successively analyzed these data sets via our proposed method and discovered some interesting shared genetic patterns. Many identified variants have been confirmed by other works. To conclude, our proposed method not only possesses a better power than related methods but also provides easily interpretable results for better understanding shared genetic architectures of complex diseases/trais.

In this work, we mainly focus on the analysis of summary statistics. With the development of new technology, more and more supplementary information, such as functional annotation data, structural data, and biochemical data, can be quickly obtained. In the future work, we will integrate these information in our method to increase the statistical power.

\section*{Acknowledgments}
This work was supported by Georgia State University Deep Grant, Hong Kong Baptist University Strategic Development Fund, Hong Kong Baptist University grant, and Hong Kong Research grant HKBU12202114.

\newpage

\title{\textbf{
\begin{center}
Supplementary Document for ``Exploring the genetic patterns of complex diseases via the integrative genome-wide approach"
\end{center}
}}

\maketitle

\section*{List}

\begin{itemize}
\item Table S1: 32 GWASs data.
\item Table S2: 32 GWASs data.
\item Table S3: Results of four methods when simulations are generated from pattern 1 with different SNRs.
\item Table S4: Results of four methods when simulations are generated from pattern 2 with different SNRs.
\item Table S5: Results of four methods when simulations are generated from pattern 3 with different SNRs.
\item Table S6: Results of four methods when simulations are generated from pattern 4 with different SNRs.
\end{itemize}
\newpage

\section*{Data descriptions}
We applied our method to analyze 32 independent diseases/traits, including

\begin{itemize}

\item 3 anthropometrics related data: body mass index \cite{speliotes2010association}, height \cite{allen2010hundreds}, waist-hip ratio adjusted for BMI \cite{heid2010meta}.

\item 9 pyschiatry related data: five PGC data \cite{cross2013identification} (attention-deficit/hyperactivity disorder, autism spectrum disorder, bipolar disorder, major depressive disorder, schizophrenia) and four TAG data \cite{tobacco2010genome} (TagCPD, TagEVRSMK, TagFORMER, TagLOGONSET).

\item 8 CAD data: total cholesterol \cite{teslovich2010biological}, low density lipoprotein \cite{teslovich2010biological}, triglycerides \cite{teslovich2010biological}, high density lipoprotein \cite{teslovich2010biological}, type 2 diabetes \cite{morris2012large}, coronary artery disease \cite{deloukas2013large}, diastolic blood pressure \cite{international2011genetic}, systolic blood pressure \cite{international2011genetic}.

\item 2 social science related data \cite{rietveld2013gwas}: edu\_years, college.

\item 2 glycaemic traits \cite{manning2012genome}: fasting glucose, fasting insulin.

\item 7 inflammatory bowel disease data: crohn's disease \cite{jostins2012host}, multiple sclerosis \cite{fingerprinting2007risk}, psoriasis \cite{feng2009multiple}, rheumatoid arthritis \cite{stahl2010genome}, type 1 diabetes \cite{barrett2009genome}, ulcerative colitis \cite{anderson2011meta}.
\item  systemic lupus erythematosus \cite{hom2008association}.
\item  parkinson \cite{simon2009genome}.
\end{itemize}

\begin{table}[H]
\centering
\caption[\textbf{32 GWASs data.}]{\textbf{32 GWASs data.} }
\begin{tabular}{p{3.5cm}|p{1.7cm}|p{11cm}}
\hline
\textbf{Name}  & \textbf{\# of SNPs} &\textbf{Link }\\ \hline

 body mass index \cite{speliotes2010association} &2471516  &	\url{http://www.broadinstitute.org/collaboration/giant/index.php/} \\ \hline
 height	\cite{allen2010hundreds}&2469635	&	\url{http://www.broadinstitute.org/collaboration/giant/index.php/}\\ \hline
 crohn's disease \cite{jostins2012host}&	953241		&	\url{http://www.ibdgenetics.org/downloads.html }\\ \hline
 fasting glucose \cite{manning2012genome} &	2628879		&\url{http://www.magicinvestigators.org/downloads/} \\ \hline
 total cholesterol \cite{teslovich2010biological}	&2693413	&\url{http://www.sph.umich.edu/csg/abecasis/public/lipids2010/}\\ \hline
 low density lipoprotein  \cite{teslovich2010biological}&	2692564	&	\url{http://www.sph.umich.edu/csg/abecasis/public/lipids2010/} \\ \hline
 triglycerides \cite{teslovich2010biological}&2692560 \cite{teslovich2010biological} &	\url{http://www.sph.umich.edu/csg/abecasis/public/lipids2010/} \\ \hline
 high density lipoprotein \cite{teslovich2010biological}&	2692429	&	\url{http://www.sph.umich.edu/csg/abecasis/public/lipids2010/} \\ \hline
 coronary artery disease \cite{deloukas2013large} 	&2420360		&\url {http://www.cardiogramplusc4d.org/downloads/}\\ \hline
 college \cite{rietveld2013gwas}	&2321510		&	\url{http://ssgac.org/Data.php}\\ \hline
 diastolic blood pressure \cite{international2011genetic}	&2461325	&			\url{http://www.ncbi.nlm.nih.gov/projects/gap/cgi-bin/study.cgi?study\_id=phs000585.v1.p1} \\ \hline
 systolic blood pressure \cite{international2011genetic}	&2461325	&			\url{http://www.ncbi.nlm.nih.gov/projects/gap/cgi-bin/study.cgi?study\_id=phs000585.v1.p1} \\ \hline
 eduyears \cite{rietveld2013gwas}&	2310087		&	\url{http://ssgac.org/Data.php} \\ \hline
 fasting Insulin \cite{manning2012genome}	&2627848		&	\url{http://www.magicinvestigators.org/downloads/}\\ \hline
 multiple sclerosis \cite{fingerprinting2007risk}	&327094		&	\url{http://www.ncbi.nlm.nih.gov/projects/gap/cgi-bin/analysis.cgi?study\_id=phs000139.v1.p1\&phv=65549\&phd=1061\&pha=2854\&pht=621\&phvf=\&phdf=\&phaf=\&phtf=\&dssp=1\&consent=\&temp=1}
  \\\hline
 parkinson \cite{simon2009genome}	&453217	&			 \url{http://www.ncbi.nlm.nih.gov/projects/gap/cgi-bin/analysis.cgi\?study\_id=phs000089.v3.p2&phv=24040&phd=392&pha=2868&pht=178&phvf=&phdf=&phaf=&phtf=&dssp=1&consent=&temp=1}
 \\ \hline

\end{tabular}
\end{table}

\begin{table}[H]
\centering
\caption[\textbf{32 GWASs data.}]{\textbf{32 GWASs data.} }
\begin{tabular}{p{3.5cm}|p{1.7cm}|p{11cm}}
\hline
\textbf{Name}  & \textbf{\# of SNPs} &\textbf{Link }\\ \hline
 attention-deficit/hyperactivity disorder \cite{cross2013identification}	&1219805		& \url{http://www.med.unc.edu/pgc/downloads}\\ \hline
 autism spectrum disorder \cite{cross2013identification}	&1219805	&	\url{http://www.med.unc.edu/pgc/downloads}\\ \hline
 bipolar disorder \cite{cross2013identification}	&1219805	&	\url{http://www.med.unc.edu/pgc/downloads}\\ \hline
 major depressive disorder \cite{cross2013identification}	&1219805	& \url{http://www.med.unc.edu/pgc/downloads}\\ \hline
 schizophrenia \cite{cross2013identification}	&1219805	&	\url{http://www.med.unc.edu/pgc/downloads}\\ \hline
 psoriasis \cite{feng2009multiple}	&440153		&		\url{http://www.ncbi.nlm.nih.gov/projects/gap/cgi-bin/analysis.cgi\?study_id=phs000019.v1.p1\&phv=20012\&phd=179\&pha=2855\&pht=63\&phvf=\&phdf=\&phaf=\&phtf=\&dssp=1\&consent=\&temp=1}
 \\ \hline
 rheumatoid arthritis \cite{stahl2010genome}	&2556271	&		\url{http://www.broadinstitute.org/ftp/pub/rheumatoid\_arthritis/Stahl\_etal\_2010NG/}\\ \hline
 type 1 diabetes	\cite{barrett2009genome}&503181	&		\url{http://www.ncbi.nlm.nih.gov/projects/gap/cgi-bin/analysis.cgi\?study_id=phs000180.v2.p2\&phv=73462\&phd=1548\&pha=2862\&pht=789\&phvf=\&phdf=\&phaf=\&phtf=\&dssp=1\&consent=\&temp=1}
 \\ \hline
 type 2 diabetes \cite{morris2012large}	&2473441	&		\url{http://diagram-consortium.org/downloads.html}\\ \hline
 TagCPD		\cite{tobacco2010genome}&2459118	&	\url{http://www.med.unc.edu/pgc/downloads}\\ \hline
 TagEVRSMK	\cite{tobacco2010genome}&2455846		&	\url{http://www.med.unc.edu/pgc/downloads}\\ \hline
 TagFORMER	\cite{tobacco2010genome}	&2456554		&	\url{http://www.med.unc.edu/pgc/downloads}\\ \hline
 TagLOGONSET \cite{tobacco2010genome}		&2457545		&	\url{http://www.med.unc.edu/pgc/downloads}\\ \hline
 ulcerative colitis \cite{anderson2011meta} 	&1428749		&	\url{http://www.ibdgenetics.org/}\\ \hline
 waist-hip ratio adjusted for BMI \cite{heid2010meta}	&2483326		&\url{http://www.broadinstitute.org/collaboration/giant/index.php/GIANT\_consortium\_data\_files}\\ \hline
 systemic lupus erythematosus \cite{hom2008association} &258402	&\url{http://www.ncbi.nlm.nih.gov/projects/gap/cgi-bin/analysis.cgi?study_id=phs000122.v1.p1&phv=66336&phd=&pha=2848&pht=629&phvf=&phdf=&phaf=&phtf=&dssp=1&consent=&temp=1}\\ \hline

\end{tabular}
\end{table}

\begin{table*}[htbp]\footnotesize
\centering
\caption{{\footnotesize\textbf{Results of four methods when simulations are generated from pattern 1 with different SNRs.} }}
\begin{tabular}{|c|c|c|c|c|c|c|c|c|c|c|c|c|}
\hline
{} &
\multicolumn{4}{c|}{SNR = $2.5$} &
\multicolumn{4}{c|}{SNR = $2.1$} &
\multicolumn{4}{c|}{SNR = $1.7$}  \\
\cline{2-13}
  & Sparsebc & SSVD & LAS & Low-rank & Sparsebc & SSVD & LAS & Low-rank & Sparsebc & SSVD & LAS & Low-rank\\
\hline

Precision & 0.95 & 0.95& 1 &0.84 &0.88 &0.94 &1 &0.75 &0.71 &0.92 &1 &0.61\\
\hline

Recall &  0.95 &0.99 &0.19 &0.82 &0.96 &0.94 &0.17 &0.82 & 0.70 & 0.76& 0.15 &0.82\\
\hline

F1-score & 0.95 &0.96 &0.32 &0.83 & 0.92 &0.94 &0.29 &0.78& 0.70& 0.82&0.26 &0.70\\
\hline
\end{tabular}
\end{table*}

\begin{table*}[htbp]\footnotesize
\centering
\caption{{\footnotesize\textbf{Results of four methods when simulations are generated from pattern 2 with different SNRs.} }}
\begin{tabular}{|c|c|c|c|c|c|c|c|c|c|c|c|c|}
\hline
{} &
\multicolumn{4}{c|}{SNR = $3.3$} &
\multicolumn{4}{c|}{SNR = $2.8$} &
\multicolumn{4}{c|}{SNR = $2.2$}  \\
\cline{2-13}
  & Sparsebc & SSVD & LAS & Low-rank & Sparsebc & SSVD & LAS & Low-rank & Sparsebc & SSVD & LAS & Low-rank\\
\hline

Precision & 0.72 &0.80 & 0.98 &0.85 &0.69 &0.87 &0.97 &0.75 &0.72 &0.80 &1 &0.69\\
\hline

Recall &  0.68 &0.80 &0.27 &0.85 & 0.57 &0.72 &0.22 &0.86 & 0.46 &0.63 &0.18 &0.74\\
\hline

F1-score & 0.69 &0.80 &0.42 &0.85 & 0.61 &0.79 &0.36 &0.80& 0.56 &0.71 &0.30 &0.71\\
\hline
\end{tabular}
\end{table*}

\begin{table*}[htbp]\footnotesize
\centering
\caption{{\footnotesize\textbf{Results of four methods when simulations are generated from pattern 3 with different SNRs.} }}
\begin{tabular}{|c|c|c|c|c|c|c|c|c|c|c|c|c|}
\hline
{} &
\multicolumn{4}{c|}{SNR = $2.6$} &
\multicolumn{4}{c|}{SNR = $2.2$} &
\multicolumn{4}{c|}{SNR = $1.8$}  \\
\cline{2-13}
  & Sparsebc & SSVD & LAS & Low-rank & Sparsebc & SSVD & LAS & Low-rank & Sparsebc & SSVD & LAS & Low-rank\\
\hline

Precision & 0.99 &0.74 &1 &0.84 &0.85 &0.74 &1 &0.75 &0.87 &0.78 &1 &0.77\\
\hline

Recall &  0.61 &0.79 &0.29 &0.86&0.63 &0.79& 0.22 &0.84& 0.56 &0.69 &0.22 &0.75\\
\hline

F1-score & 0.76 &0.77 &0.45 &0.85&0.73 &0.76 &0.37 &0.79  &0.68 &0.73 &0.37 &0.76\\
\hline
\end{tabular}
\end{table*}

\begin{table*}[htbp]\footnotesize
\centering
\caption{{\footnotesize\textbf{Results of four methods when simulations are generated from pattern 4 with different SNRs.} }}
\begin{tabular}{|c|c|c|c|c|c|c|c|c|c|c|c|c|}
\hline
{} &
\multicolumn{4}{c|}{SNR = $2.9$} &
\multicolumn{4}{c|}{SNR = $2.4$} &
\multicolumn{4}{c|}{SNR = $1.9$}  \\
\cline{2-13}
  & Sparsebc & SSVD & LAS & Low-rank & Sparsebc & SSVD & LAS & Low-rank & Sparsebc & SSVD & LAS & Low-rank\\
\hline

Precision & 0.86 &0.76 & 0.99 &0.80 &0.81 &0.85 &1 &0.78 &0.75 &0.86 &1 &0.71\\
\hline

Recall &  0.64 & 0.68 &0.27 &0.83& 0.61 &0.48 &0.20 &0.76& 0.50 &0.48 &0.19 &0.71\\
\hline

F1-score & 0.72 &0.72& 0.43 &0.82& 0.69 &0.62 &0.23 &0.77 & 0.59& 0.62& 0.32& 0.71\\
\hline
\end{tabular}
\end{table*}

\bibliographystyle{natbib}     
\bibliography{mybib}     

\end{document}